%% file: main.tex
\documentclass[conference]{IEEEtran}
\IEEEoverridecommandlockouts

\input{preamble}

\begin{document}

\input{header}

\input{paper}

\clearpage

\bibliographystyle{IEEEtran}
\bibliography{citations}

\end{document}

%% file: preamble.tex
\usepackage{amssymb}
\usepackage{amsmath}
\usepackage{comment}
\usepackage[draft]{hyperref}
\usepackage{cleveref}
\usepackage{graphicx}
\usepackage{siunitx}
\usepackage{enumitem}
\usepackage{subfloat}

\usepackage{graphicx}
\usepackage[caption=false]{subfig}

\usepackage{xcolor} 
\usepackage{url}
\usepackage{hyperref} 
\usepackage{cleveref} 

\usepackage{algorithm} 
\usepackage[noend]{algpseudocode}

\usepackage{orcidlink}

\newcommand{\FULLVERSION}{}

%% file: header.tex
\title{Ethereum Conflicts Graphed
\Large}

\newcommand{\AuthorNameA}{Dvir David Biton}
\newcommand{\AuthorEmailA}{dbiton@campus.technion.ac.il}
\newcommand{\AuthorORCIDA}{0009-0009-3440-375X}

\newcommand{\AuthorNameC}{Yaron Hay}
\newcommand{\AuthorEmailC}{yaron.hay@cs.technion.ac.il}
\newcommand{\AuthorORCIDC}{0009-0006-1263-7318}

\newcommand{\AuthorNameB}{Roy Friedman}
\newcommand{\AuthorEmailB}{roy@cs.technion.ac.il}
\newcommand{\AuthorORCIDB}{0000-0001-6460-9665}

\newcommand{\AuthorInstituion}{Technion}
\newcommand{\AuthorDepartment}{Computer Science}
\newcommand{\AuthorCity}{Haifa}
\newcommand{\AuthorCountry}{Israel}

\author{%
 \IEEEauthorblockN{\AuthorNameA}
 \IEEEauthorblockA{\textit{\AuthorDepartment} \\
 \textit{\AuthorInstituion},
 \AuthorCity, \AuthorCountry \\
 \orcidlinkc{\AuthorORCIDA}\\
 \href{mailto:\AuthorEmailA}{\AuthorEmailA}
 }
 \and
 \IEEEauthorblockN{\AuthorNameB}
 \IEEEauthorblockA{\textit{\AuthorDepartment} \\
 \textit{\AuthorInstituion},
 \AuthorCity, \AuthorCountry \\
 \orcidlinkc{\AuthorORCIDB}\\
 \href{mailto:\AuthorEmailA}{\AuthorEmailB}
 }
 \and
 \IEEEauthorblockN{\AuthorNameC}
 \IEEEauthorblockA{\textit{\AuthorDepartment} \\
 \textit{\AuthorInstituion}, \AuthorCity, \AuthorCountry \\
 \orcidlinkc{\AuthorORCIDC}\\
 \href{mailto:\AuthorEmailA}{\AuthorEmailC}
 }
}

\maketitle
\begin{abstract}
Ethereum, a leading blockchain platform, has revolutionized the digital economy by enabling decentralized transactions and the execution of smart contracts. Ethereum transactions form the backbone of its network, facilitating peer-to-peer exchanges and interactions with complex decentralized applications. Smart contracts extend Ethereum's capabilities by automating processes and enabling trustless execution of agreements.
Hence, understanding how these smart contracts interact is important in order to facilitate various performance optimizations, such as warming objects before they are being accessed and enabling concurrent execution.
Of particular interest to us are the development of the calling graph, as well as the read sets and write sets of invocations within the same block, and the properties of the associated conflict graph that is derived from them.
The latter is important for understanding the parallelization potential of smart contracts on Ethereum. We traced upwards of 2 million recent Ethereum blocks using call tracer and prestate tracer, out of a total of 21.4 million blocks at the time of writing. We report on the transactions per block distribution, the structure of call trees in smart contract invocations, the ratio of value-transfer transactions to smart contract invocations, as well as provide a comprehensive study of the structure of blocks' conflict graphs. We find that conflict graphs predominantly show a star like configuration, as well as other noteworthy structural properties.
The data~\cite{EthStats} and code~\cite{Github} are available in open source.

\end{abstract}


%% file: paper.tex
\section{Introduction} \label{sec:introduction}



The rise of blockchain technology is part of a transition to decentralized systems, especially in the financial world. Smart contract transactions provide the means for developing new applications that benefit from blockchains’ properties; therefore, they are key for connecting blockchains to new markets. The execution of all transactions within a block must be deterministically serializable~\cite{Schneider1990} to ensure correct replication among all replicas, including any smart contract transactions. That is, all executions of the transactions must be equivalent to the same sequential execution. Further, the execution of smart contract transactions is known to be the main limitation on blockchains’ throughput~\cite{diablo}.

Parallel transaction processing has been suggested for executing smart contracts \cite{ParBlockchain,ConcurrencySC,BlockSTM,KD04,EmpSdy-Con}. Most of these efforts have focused on executing the transactions according to a predetermined logical order. Thus, the execution of conflicting transactions must respect the logical order to prevent inconsistent results. Some works \cite{ParBlockchain,KD04,Blochchain-DB,ColoringSC} have focused on the undirected conflict graph which connects each pair of conflicting transactions with an edge; they direct the edges of the conflict graph to form a DAG and execute the transactions according to it. The work of \cite{ColoringSC} has suggested optimizing the order of transactions according to the aggregated execution time of a block’s transactions.

These techniques are sensitive to the characteristics of the conflict graph. 
Alas, most evaluations of parallel execution techniques rely on synthetic transactions such as randomized balance transfers~\cite{ParBlockchain,ConcurrencySC}, key-value operations~\cite{YCSB}, database workloads such as TPC-C~\cite{tpc-c} and TPC-E~\cite{tpc-e}, while some remain purely theoretical~\cite{ColoringSC}.
Even the recently proposed Diablo~\cite{diablo} benchmark does not provide any information on conflicts between transactions.
The difference between such workloads and practical blockchain workloads doubts the implications of these techniques for practical use in blockchains, particularly in the absence of smart contract transactions.  

In this paper, we study the intra-block conflict graphs associated with transactions on Ethereum~\cite{ethereum}, currently the most popular smart contract blockchain. 
To that end, we examine more than 2M recent blocks from the Ethereum mainnet, and extract statistics about various characteristics of the conflicts graphs associated with these blocks\ifdefined\FULLVERSION, focusing on read-write conflicts\fi.
We start by methodically identifying the read-sets and write-sets of Ethereum transactions, not a trivial task by itself.
This, in turn, enables us to derive the conflict graph for each block.
Properties of the resulting conflict graphs are then gathered and analyzed.
We are particularly interested in their implications on the potential for concurrent execution of~transactions.
We also track the calling trees of smart contracts, as well as the various smart contracts'~popularity.

\paragraph*{Summary of Findings}
Briefly, we find that block sizes typically center around 150 transactions, although they can reach up to 1,400, and occasionally 0. 
Call trees tend to be narrow, averaging a height of 4 with around 16 leaves, which helps control gas costs by avoiding extensive nesting.
About 60\% of transactions in each block are pure value transfers, though this fraction varies significantly.

\ifdefined\FULLVERSION
We analyze the distribution of conflicts between read-write and write-read vs. all conflicts, including write-write conflicts.
We discover that when accounting also the write-write conflicts, nearly all transactions conflict with each other, whereas when considering only read-write and write-read conflicts, much more interesting structures appear.
It is well known that when using a multi-versioned data store, it is possible to maintain serializability without tracking and preventing write-write conflicts~\cite{ParBlockchain}.
Hence, we focus almost entirely on read-write and write-read conflicts, and from now on, unless explicitly mentioned otherwise, when saying conflicts we refer to read-write and write-read only.
Yet, we also analyze the main sources of write-write conflicts, and discover that they are caused by a small number of contracts and a small number of shared resources, typically infrastructure wallets.
\fi

The conflict graphs generally exhibit a star-like topology, with a single hub-and-spoke structure surrounded by smaller, often single-node components.
The mean degree grows with graph density, while the maximum degree stays fairly
constant, indicating that new edges form primarily among leaf nodes rather than at the hub. 
The diameter rarely exceeds $5$, and the chromatic number remains in the range of $2$ to $5$, suggesting strong potential for parallel execution~\cite{ColoringSC}. 
Larger, denser blocks can have long simple paths exceeding $500$ nodes, resulting in a ratio of longest path to chromatic number reaching up to~$100$.
In the paper, we elaborate on these results and their significance for parallelizing Ethereum smart contracts~execution.

\paragraph*{Paper Roadmap} In \Cref{sec:related} we survey related work. \Cref{sec:prelim} discusses preliminary definitions regarding conflict graphs. \Cref{sec:ether} presents the abstraction Ethereum introduces for smart contracts, and \Cref{sec:method} discusses the methodology we use to construct and analyze the conflict graphs based on Ethereum blocks. In \Cref{sec:results} we summarize the main statistics, and we conclude with a discussion in \Cref{sec:discussion}.

\section{Related Work}
\label{sec:related}

We are unaware of other works that have explored the conflict graphs that correspond to smart contracts execution in Ethereum.
While the EIP-2930 transaction type was introduced into Ethereum in order to enable declaring what objects would be accessed by a transaction, the work of~\cite{Dissecting-EIP-2930} has found that it is used quite rarely despite giving discounted gas prices.
Due to the potential for system optimizations, such as pre-warming of objects and the potential for parallelism, as highlighted in our findings, it seems that the Ethereum community should do more to encourage its usage.

Speculative parallelizing smart contracts execution on the Ethereum VM was proposed and tested in~\cite{VH20}.
In such a method, all transactions of the same block are first executed concurrently.
At commit time, transactions that conflicted with previously committed transactions are aborted and get re-executed.
They examined this method on sampled blocks from 2016 and 2017, and have found that the potential for speedup using this technique has dropped from 8 fold in 2016 to only 2 in 2017, due to the rise in conflicts which caused many more aborts.
They also discovered that a few smart contracts were responsible for most conflicts.

Several academic works have explored parallelizing transactions execution on blockchains~\cite{ParBlockchain,Blochchain-DB} and Byzantine tolerant SMRs in general~\cite{KD04,COS}.
Indeed, a host of contemporary blockchains these days employ parallelization, e.g.,~\cite{BlockSTM,solana-sealevel,sui,sei-2,monad}.
Most of them rely on having the transactions declare its read sets and write sets a-priori, but some use optimistic ordered execution to avoid this need~\cite{BlockSTM}.
In all these works, whenever two transactions conflict, their commit order must obey the order in which they appeared in the block.
In contract, the work of~\cite{ColoringSC} discovered that coloring the conflict graph and then ordering conflicting transactions in accordance to their respective color number can yield significant performance gains, proportional to the ratio between the chromatic number of the conflict graph and the longest simple path in the graph.
In this study, we find that this is the case in Ethereum.

The evaluation of deterministic concurrency controls for blockchain is mostly done via synthetic workloads such as randomized P2P balance transfers~\cite{ParBlockchain}, randomized P2P multiple input-output transfer transactions~\cite{BlockSTM}, and synthetic Smart Contracts~\cite{ConcurrencySC}. The closely related topic of deterministic databases \cite{DetOverview,Aria,caracal} uses standard benchmarks like TPC-C \cite{tpc-c}, TPC-E~\cite{tpc-e}, and YCSB~\cite{YCSB}. The Diablo Blockchain Benchmark~\cite{diablo} is designed to benchmark raw computational power rather than mimic real accesses and conflicts.

There have been multiple studies of various aspects of the Ethereum network over the years, e.g.,~\cite{EthPerfAnalysis,EthInfoProp,EthGasSize,ETHLargeBlock,XBlock-ETH,RoleReward}.
However, to the best of our knowledge, none of them explored the conflicts graphs associated with smart contracts transactions.

\section{Preliminary}
\label{sec:prelim}
We assume that an arbitrary block $B$ consists of a set $T$ of $n$ transactions. Each transaction is an atomic program with access to the blockchain's objects; transactions may access them for both reading and writing purposes.

\subsection{Conflict Graphs}
Given an arbitrary transaction $tx \in T$, we denote its \emph{read set} as the set of all objects that $tx$ may read from during its execution. Similarly, we denote its \emph{write set} as all objects $tx$ may write to.
Following traditional database terminology, we say that two transactions $tx_1\in T$ and $tx_2\in T$ are \emph{conflicting} if they have a shared object such that both $tx_1$ and $tx_2$ write to it \ifdefined\FULLVERSION (write-write conflict) \fi or read from it and write to it\ifdefined\FULLVERSION \ (write-read or read-write conflict)\fi.
\ifdefined\FULLVERSION
Recall that unless explicitly mentioned otherwise, in this paper we focus only on conflicts where one of the operations reads from the shared object.
\fi
We construct the \emph{conflict graph} for the block $B$ by connecting each pair of conflicting transactions with an undirected edge.

\subsection{Graph Properties}


Graphs can be characterized by multiple properties. 
In this work we focus on the following properties of conflict graphs to better understand their characteristics when derived from blocks of the Ethereum~blockchain.

\begin{LaTeXdescription} 
\item[Density:] the ratio of the number of edges to the maximum possible edges in an undirected graph.
\item[Diameter:] The longest of the shortest paths between each pair of nodes in the graph. 
\item[Chromatic number:] The minimum number of colors required to color the nodes of the graph so that no two adjacent nodes share the same color. 
\item[Clique number:] The size of the largest complete subgraph (clique) in the graph, where every pair of nodes is directly connected by an edge. This serves as a lower bound for the chromatic number.
\item[Longest simple path:] The length of the longest path in the graph that does not revisit any node.
\item[Largest connected component:] The size of the largest subset of nodes such that there exists a path between any two nodes in this subset. This serves as an upper bound for the longest simple path.
\end{LaTeXdescription}

\section{Ethereum}
\label{sec:ether}

Ethereum operates as a decentralized platform that facilitates peer-to-peer transactions while enabling the execution of smart contracts. To fully appreciate Ethereum's functionality and statistical analysis, it is essential to understand the foundational concepts of transactions, smart contracts, and the challenges posed by transaction conflicts within the network.

\subsection{Transactions}
An Ethereum transaction represents a message sent from one account to another. Transactions can involve transferring Ether (ETH), the network's native cryptocurrency, or interacting with smart contracts deployed on the Ethereum Virtual Machine (EVM). Each such transaction includes the following~components:
\begin{LaTeXdescription} 
\item[Sender and Receiver Addresses:] Identifiers of the accounts involved in the transaction. Both smart contracts and individual wallets are accounts with an Ethereum address.
\item[Transaction Value:] The amount of ETH being transferred.
\item[Data Field:] Optional data used to trigger specific smart contract functions or carry arbitrary information.
\item[Gas and Gas Price:] Metrics defining the computational resources required for the transaction and the price the sender is willing to pay per unit of gas.
\item[Nonce:] A unique value used to prevent transaction replay attacks and ensure transaction order.
\end{LaTeXdescription}
Once initiated, transactions are propagated across the network, validated by nodes, and eventually included in a block by miners or validators, depending on the consensus mechanism. 

\subsection{Smart Contracts}
Smart contracts are self-executing code stored on the Ethereum blockchain. They enable automated and decentralized interactions between parties without requiring a central authority or intermediary. Smart contracts can implement complex logic, enforce conditions, and manage state across the blockchain. They implement many of Ethereum's most notable use cases, including decentralized finance (DeFi), non-fungible tokens (NFTs), and supply chain solutions.

\subsection{Tools}

\subsubsection{Call Tracer}

The \emph{callTracer}~\cite{geth_tracing} records all call frames executed during a transaction, including the top-level call (depth 0). It returns a nested list of call frames forming a tree: the root is the top-level call, and sub-calls are its children. Each call frame includes the fields listed in \Cref{table:calltracer}.

\begin{table}[!h]
\centering
\begin{tabular}{|c c|}
\hline
\textbf{Field} & \textbf{Description} \\
\hline
\texttt{type}          & Call type (e.g., CALL, DELEGATECALL) \\
\hline
\texttt{from}          & Originating address \\
\hline
\texttt{to}            & Target address \\
\hline
\texttt{value}         & Value transferred (in Wei) \\
\hline
\texttt{gas}           & Gas provided for the call \\
\hline
\texttt{gasUsed}       & Gas consumed by the call \\
\hline
\texttt{input}         & Call input data \\
\hline
\texttt{output}        & Return data from the call \\
\hline
\texttt{error}         & Error, if any occurred \\
\hline
\texttt{revertReason}  & Solidity revert reason (if revert occurred) \\
\hline
\texttt{calls}         & List of sub-calls \\
\hline
\end{tabular}

\caption{Fields returned by the call tracer for each call frame.}
\label{table:calltracer}
\end{table}

Each call frame may read from or write to the \texttt{from} or \texttt{to} address. The call type determines its permissions. These permissions apply not only to the call itself but also propagate to any sub-calls it invokes. For example, a \texttt{STATICCALL} enforces a read-only context, disallowing state modifications. Thus, a \texttt{STATICCALL} cannot create new contracts via \texttt{CREATE}, as that would involve state modifications. Similarly, if a \texttt{CALL} is made within a \texttt{STATICCALL} context, the \texttt{CALL} will inherit the read-only restriction and be unable to modify~state.

This cascading rule ensures that a \texttt{STATICCALL} context remains truly immutable throughout all its sub-calls, providing strong guarantees of immutability for execution reliant on \texttt{STATICCALL} semantics.

\subsubsection{Prestate Tracer}

The \emph{prestate tracer}~\cite{geth_tracing} records changes to the state caused by a transaction. It has two modes, controlled by the \texttt{diffMode} parameter:

\begin{itemize}
\item \texttt{diffMode = false}: Returns the accounts and fields (nonce, code, storage) that are required to execute the transaction, indicating which addresses were read or might be~modified.
\item \texttt{diffMode = true}: Returns both the pre- and post-state for all accessed addresses and state fields, effectively showing which parts of the state changed due to the~transaction.
\end{itemize}

This tracer reexecutes the transaction and tracks every touched state element. This is similar to a stateless witness, but does not produce cryptographic proofs; only the necessary trie leaves. The tracer outputs a dictionary keyed by addresses, where each entry contains the fields present in \Cref{table:prestatetracer}.

\begin{table}[!h]
\centering
\begin{tabular}{| c c |}
\hline
\textbf{Field} & \textbf{Description} \\ 
\hline
\texttt{balance} & Account balance in Wei \\
\hline
\texttt{nonce}   & Account nonce \\
\hline
\texttt{code}    & Account bytecode \\
\hline
\texttt{storage} & Storage slots of the account \\
\hline
\end{tabular}
\caption{Fields returned by the prestate tracer for each address.}
\label{table:prestatetracer}
\end{table}

\section{Methodology}
\label{sec:method}

\subsection{Ethereum Archive Node}

We have rented an Ethereum archive node to collect traces block by block. Each block is fully simulated and traced using both the call tracer and the prestate tracer~\cite{geth_tracing}. Data were gathered using the \texttt{debug\_traceBlockByNumber} Ethereum Debug API method. We have collected call traces of transactions in 2,271,946 blocks in the range between \#19,010,000 to \#21,299,999. We also collected prestate traces, with both \texttt{diffMode}=true and false for transactions in 2,898,175 blocks in the range between \#18,261,000 to \#21,299,999.

\subsection{Read Sets, Write Sets \& Conflicts}

For the call tracer approach, we infer read/write sets from the call types and their permissions. Each call has 'to' and 'from' addresses, which can refer to a wallet or to a contract. Every call that may write to an address, we consider it as writing to it; every call that may read from an address, we consider as reading from it. This approach yields a broader (more conservative) set of possible reads and~writes. After determining each transaction’s read and write sets within a block, we construct the conflict graph by representing each transaction as a node. An edge is added between two nodes if an address appears in the write set of one transaction and in the read or write set of the other.

For the prestate tracer approach, we precisely identify the address of each modified wallet or contract by setting \texttt{diffMode = true}, which is write set. We can find the address of each wallet or contract accessed by the transaction by using \texttt{diffMode = false} - yielding the union of the read set and the write set. We find the read set as the difference of the two sets we gather. After acquiring the read and write sets of each transaction, constructing the conflict graph is done in the same method.

Although the prestate tracer offers sub-address–level detail, for our analysis we treat each address as a single, indivisible unit of potential modification.
The prestate tracer method is more accurate and less conservative than the call tracer method, which can overestimate the read/write sets.

\begin{figure}[h]
\centering
\includegraphics[width=0.48\columnwidth]{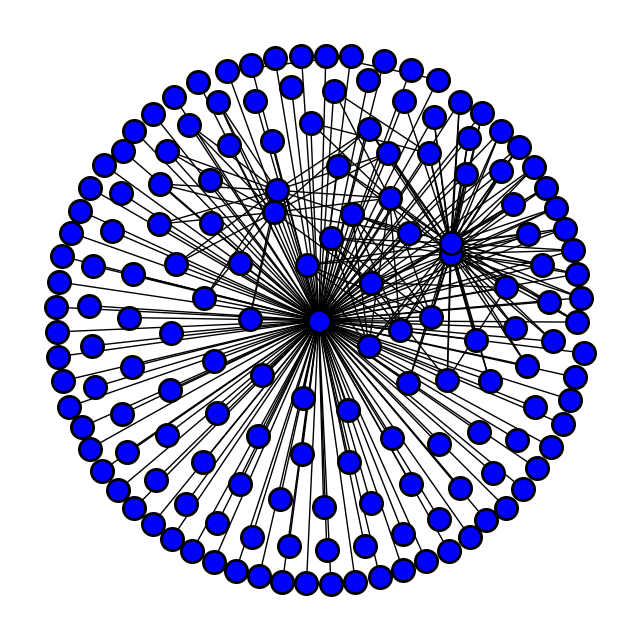}
\hfill
\includegraphics[width=0.45\columnwidth]{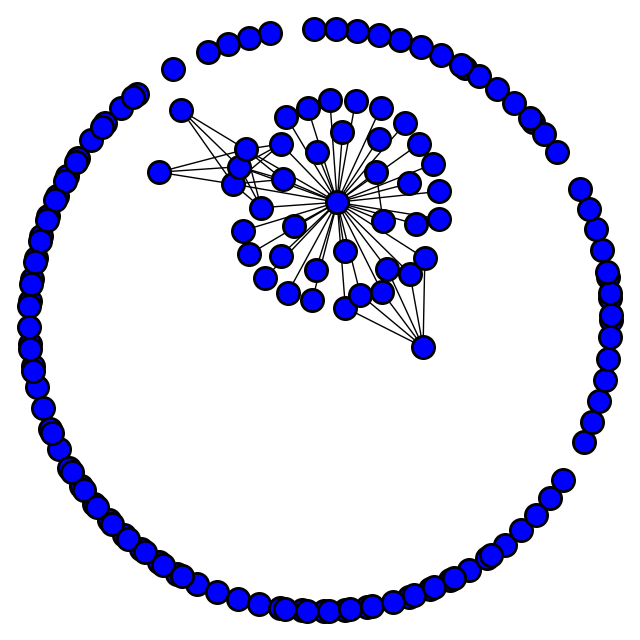}
\caption{Prestate conflict graphs of block 20700000 with density of 0.021 (left), and block 20334250 with density of 0.005 (right), the median density is 0.017}
\label{fig:conflict_graph_example}
\end{figure}

As an example, \Cref{fig:conflict_graph_example} presents the conflict graph created by prestate tracer for blocks \#20,700,000 and \#20,334,250. Each node is a transaction, and an edge indicates a conflict. The 74th transaction in block \#20,700,000 has the most conflicts by a large margin; it is the middle node in the figure. The transaction is made by "jaredfromsubway.eth" to an MEV bot that exploits vulnerabilities in DeFi protocols for profit. Note that every single transaction is in conflict in this block, creating a singular connected component. Yet, this is not always the case. Block \#20,334,250, also included in the figure, has a much lower density and includes many transactions with no conflicts at all.

\subsection{Implementation}

The collection, retrieval and processing of traces is implemented in Python 3.11, using \texttt{networkx} 3.4.2 to process conflict graphs, and \texttt{requests} to interact with the Ethereum node. Traces are compressed and stored in the h5 format using \texttt{h5py}, requiring more than 7 Terabytes of storage. Many other Python modules such as \texttt{pandas}, \texttt{numpy} and \texttt{matplotlib} were also used. The complete implementation is available and open source~\cite{Github}.

Among the metrics we collected and display below, the following metrics required some non-trivial processing of data:

\begin{itemize}
\item \textbf{Graph Coloring:} We used the greedy coloring algorithm DSATUR~\cite{dsatur} to estimate the chromatic number of conflict graphs (exact minimum coloring is NP-hard~\cite{Karp1972}).
\item \textbf{Longest Path Estimation:} We use a Monte Carlo approach to approximate the length of the longest path in the graph. We focus on the largest connected components, randomly choosing a set of nodes as start points. From each start node, we iteratively build a simple path by uniformly selecting an unvisited neighbor at each step, continuing until no unvisited neighbors remain. Connected components with fewer nodes than the current best estimate of the longest path are skipped. This approach has been shown to yield accurate longest-path estimates for random graphs~\cite{gnp-dfs}.
\end{itemize}

\section{Results}
\label{sec:results}

Below we exhibit and summarize the main results.
Detailed metrics for individual blocks, as well as additional metrics and graphs can be found in~\cite{EthStats}. 

\begin{figure}[p]
\centerline{\includegraphics[width=\columnwidth]{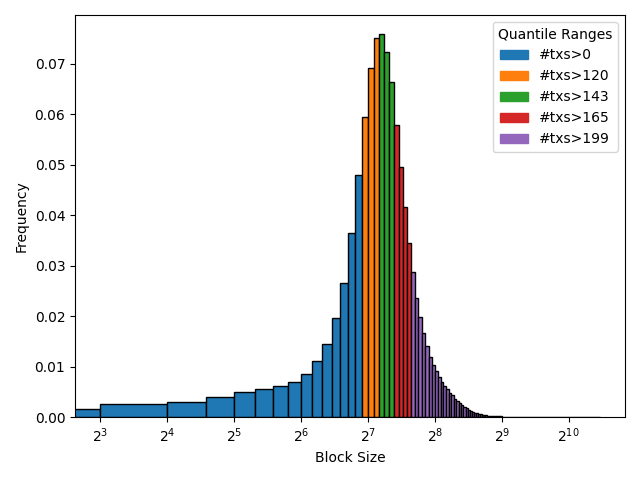}}
\caption{Block size histogram with a bucket width of 8}
\label{fig:block_size_log}
\end{figure}

\begin{figure}[p]
\centering
\includegraphics[width=\columnwidth]{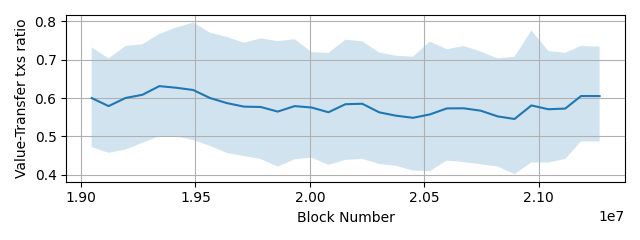}
\hfill
\includegraphics[width=\columnwidth]{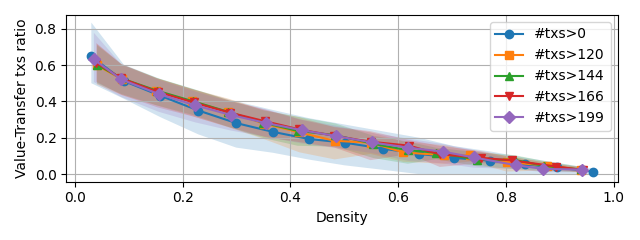}
\caption{Value-transfer count to block size ratio}
\label{fig:value_transfer_txs_ratio}
\end{figure}

\subsection{Block size}
\Cref{fig:block_size_log} plots the distribution of block sizes, i.e., the number of transactions per block, in a log-scale histogram. 
Most blocks have around 150 transactions, with a long tail reaching 1,000 transactions and more; the largest block size observed is 1,410. Very rarely, blocks have a size of 0.

\begin{figure}[htbp]
\centerline{\includegraphics[width=\columnwidth]{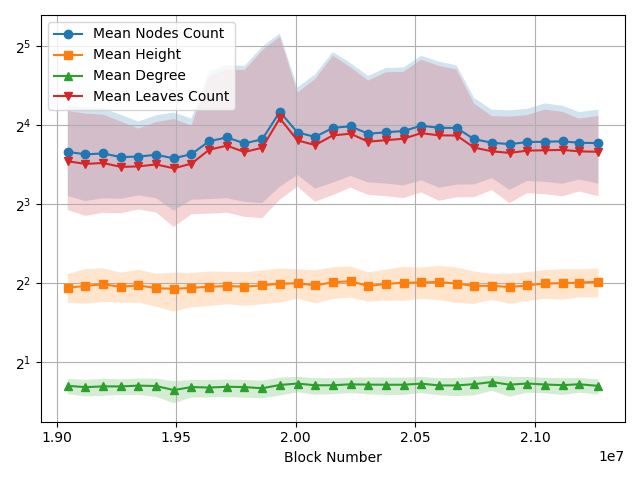}}
\caption{Smart contract call tree metrics}
\label{fig:call_metrics}
\end{figure}

\subsection{Value Transfer TXs Ratio}
In \Cref{fig:value_transfer_txs_ratio}, we plot the percentage of block transactions that are pure value transfers as functions of both block number and graph density. The average ratio hovers around 0.6 but varies considerably from block to block. By contrast, density shows a near-linear negative correlation with the percentage of value-transfer transactions—a relationship that remains largely unaffected by block size.
This indicates that dense graphs are formed when most transactions are due to smart contracts.

\subsection{Calling Trees}
\Cref{fig:call_metrics} summarizes the call metrics for blocks, reporting the mean values along with the top and bottom 5\%. Each block contains multiple transactions. For every non-value-transfer transaction, we constructed a call tree and collected four metrics: the total number of nodes, the height of the tree, the mean degree, and the number of leaves. We then computed block-level averages of these metrics.

As shown in the figure, the mean height is around 4, and the mean degree is below 2, both exhibiting low variability. In contrast, the node count fluctuates significantly, typically ranging from 8 to 32, with a mean below 16. Given that a full binary tree of height 4 has 16 nodes, most call trees are quite narrow. On average, the leaves count is slightly lower than the total node count. Combined with the small mean degree, this indicates that most nodes are leaves while a small subset of nodes may have higher degrees. Large call trees present opportunities for optimization and gas cost reduction, since nested calls tend to be relatively expensive.

\ifdefined\FULLVERSION
\begin{figure}[t]
\centerline{\includegraphics[width=\columnwidth]{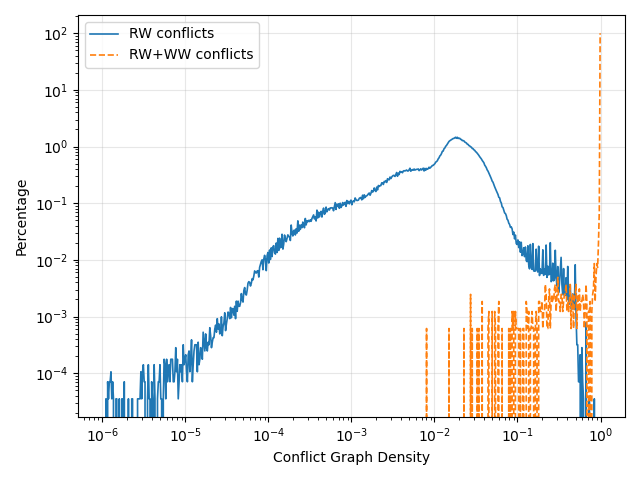}}
\caption{Density by conflict types}
\label{fig:eth-density}
\end{figure}
\fi

\ifdefined\FULLVERSION
\subsection{RW\&WW Conflicts vs RW Conflicts}

\Cref{fig:eth-density} shows the distribution of density for conflict graphs created when considering only read-write and write-read conflicts, denoted RW conflicts for short, vs. when accounting for all conflicts, denoted RW+WW conflicts. 
In addition, the call tracer over estimates conflicts more than the prestate tracer, meaning it would also have complete graphs when considering write-write conflicts.

We further analyzed the conflict graphs, checking what is the cause of the numerous write-write conflicts. We looked at the conflicts at a sub-smart contract granularity, checking whether the conflict originated from a smart contract storage slot, the nonce, modifying a wallet or modifying code. We count the number of conflicts that have occurred due to each cause in each address.

We have found that the number of unique addresses, separated by cause of conflict, is $8.68*10^6$, and the total number of WW conflicts is $4.52*10^{10}$. The top 10 most popular addresses account for $87.5\%$ of the conflicts. The distribution has an extremely long tail, with the median number of conflicts being 1, and the 95th percentile being 54. The hill estimator with k=100 is 0.74, and the mean is  5206.79.

We found that most of the write-write conflicts are caused by writes to MEV builders' wallets.
The top 10 most common sources for write-write conflicts, all due to balance conflicts,~are:

 \begin{LaTeXdescription}
  \item[Beaverbuild (\(1.72\times10^{10}\) hits):] MEV builder.
  \item[Titan Builder (\(1.12\times10^{10}\) hits):] MEV builder.
  \item[Rsync Builder (\(6.35\times10^{9}\) hits):] MEV builder.
  \item[Lido Execution-Layer Rewards Vault (\(1.28\times10^{9}\) hits):] Contract that collects priority-fee and MEV rewards for Lido stakers.
  \item[Flashbots: Builder (\(1.05\times10^{9}\) hits):] MEV builder.
  \item[Wrapped Ether (WETH) Contract (\(9.6\times10^{8}\) hits):] Canonical ERC-20 used to wrap and unwrap ETH.
  \item[jetbldr.eth (\(5.23\times10^{8}\) hits):] MEV builder.
  \item[builder0x69.eth (\(4.84\times10^{8}\) hits):] MEV builder.
  \item[Flashbots: Builder 2 (\(2.67\times10^{8}\) hits):] MEV builder.
  \item[penguinbuild.eth (\(2.39\times10^{8}\) hits):] MEV builder.
\end{LaTeXdescription}

Out of all WW conflicts observed in the approximately 3M blocks traced, 97.6\% are balance related, 1.7\% are storage related, 0.6\% are nonce related and exactly 0 conflicts were caused by code, although some WR conflicts are caused due to code modifications. Hence, the vast majority of conflicts are due to balance changes.

Following the known result that tracking and enforcing write-write conflicts is not essential for enforcing serializability when using a multi-versioned data store~\cite{ParBlockchain}, below we only explore conflict graphs created by read-write and write-read conflicts only.
\fi

\subsection{Prestate Conflict Graph}

For each block, we generated a conflict graph using the prestate tracer’s precise read/write sets. We then grouped blocks quantiles according to their transaction counts, labeling each quantile by the minimum transaction count of its constituent blocks. In the presented graphs, we plot a block-level metric, using the median value within each group to mitigate noise. The filled-in regions around the median indicate the top and bottom 5\% of observed values. We present the metric as a function of graph density, where density corresponds to the percentage of conflicting transaction pairs.

\begin{figure}
\centerline{\includegraphics[width=\columnwidth]{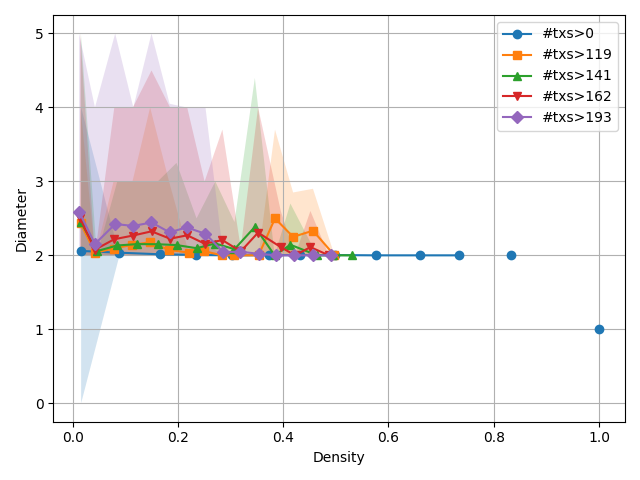}}
\caption{Prestate conflict graph diameter}
\label{fig:diameter}
\end{figure}

\begin{figure}
\centerline{\includegraphics[width=\columnwidth]{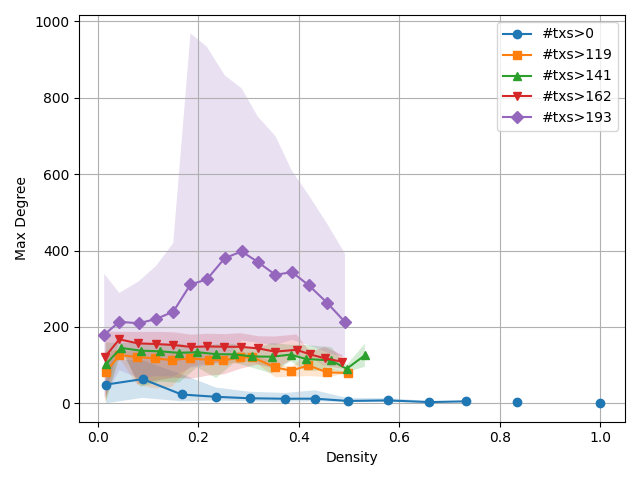}}
\caption{Prestate conflict graph max degree}
\label{fig:max_degree}
\end{figure}

\begin{figure}
\centerline{\includegraphics[width=\columnwidth]{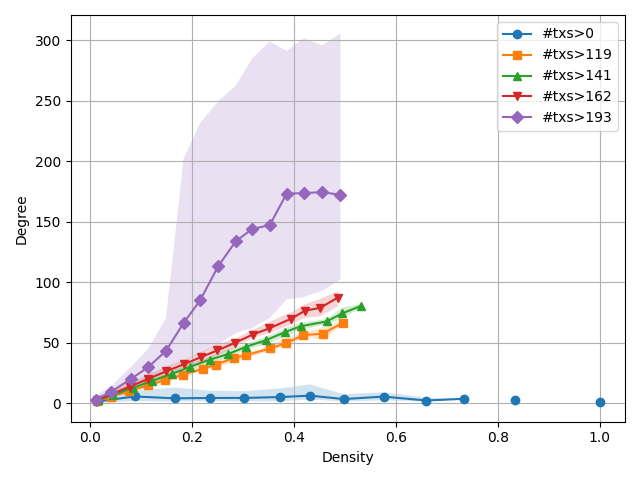}}
\caption{Prestate conflict graph mean degree}
\label{fig:degree}
\end{figure}

\begin{figure}
\centerline{\includegraphics[width=\columnwidth]{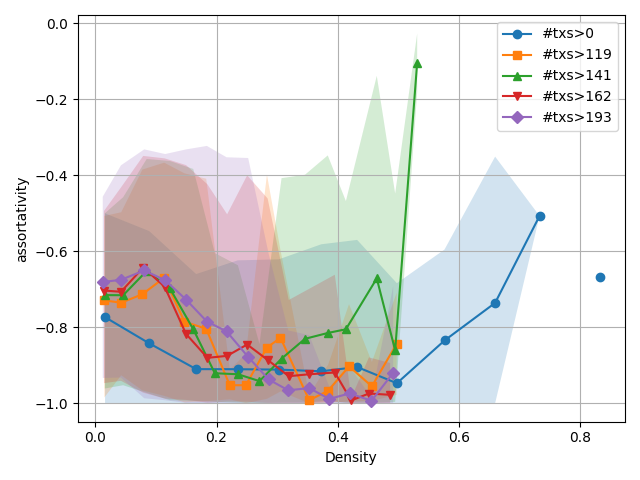}}
\caption{Prestate conflict graph assortativity}
\label{fig:assortativity}
\end{figure}

\subsubsection{Star Structure}

In many of the prestate conflict graphs we examined, a single large connected component dominates, surrounded by numerous smaller components that often consist of just one node.

The large connected component typically exhibits a star-like shape, in which a single central node is connected to a multitude of leaf nodes. In some cases, particularly in higher density graphs, these large components also contain smaller “star” substructures branching off the main~hub.
An example of such a relatively dense configuration is shown in \Cref{fig:conflict_graph_example}.

Looking at the diameter in \Cref{fig:diameter}, the median value stays consistent above 2 and rarely exceeds 5. This suggests that the structure of the graph includes connected hubs that allow short paths between nodes, rather than long chains of nodes of a small degree. In contrast, the maximum degree, shown in \Cref{fig:max_degree}, is notably larger and more sensitive to block size. The median maximum degree is typically around 100 and can reach up to 600 for larger blocks. This aligns with the star-like configurations.
That is, while the diameter remains small, due to the central hub and its immediate neighbors, the maximum degree scales roughly with block size, reflecting the growing 'hub' node's increasing number of connected leaves.

In \Cref{fig:degree}, we observe that the mean degree increases linearly with the density of the graph, except for very small and very large blocks. In contrast, as shown in \Cref{fig:max_degree}, the maximum degree does not show a clear upward trend as density increases. Instead, it remains relatively stable.

In \Cref{fig:assortativity}, negative assortativity values are consistently observed across graph densities. High-degree nodes (hubs) predominantly connect to low-degree nodes (leaves), while leaf nodes rarely interconnect. Occasional sharp fluctuations at higher densities may indicate secondary hubs forming hierarchical structures, intensifying the hub-and-spoke pattern.

This pattern suggests that most graphs feature a similarly sized star-like structure regardless of density. As the density grows, the additional edges tend not to attach to the central “hub” node of the star, but rather form between the leaf nodes themselves or between separate smaller star-like components. Thus, while the mean degree rises with density, the maximum degree stays comparatively constant, reflecting the persistent presence of a dominant hub node.

\begin{figure}
\centerline{\includegraphics[width=\columnwidth]{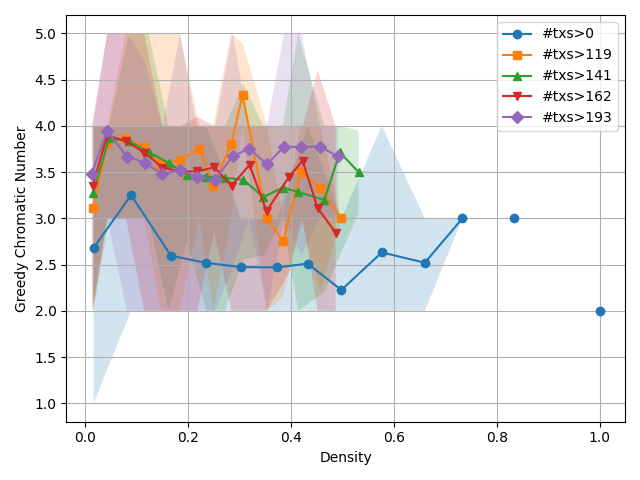}}
\caption{Prestate conflict graph greedy color}
\label{fig:greedy_color}
\end{figure}

\begin{figure}
\centerline{\includegraphics[width=\columnwidth]{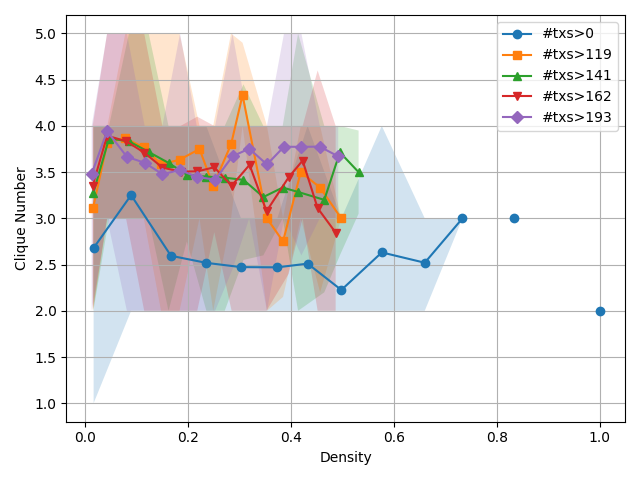}}
\caption{Prestate conflict graph clique number}
\label{fig:clique_number}
\end{figure}

\begin{figure}
\centerline{\includegraphics[width=\columnwidth]{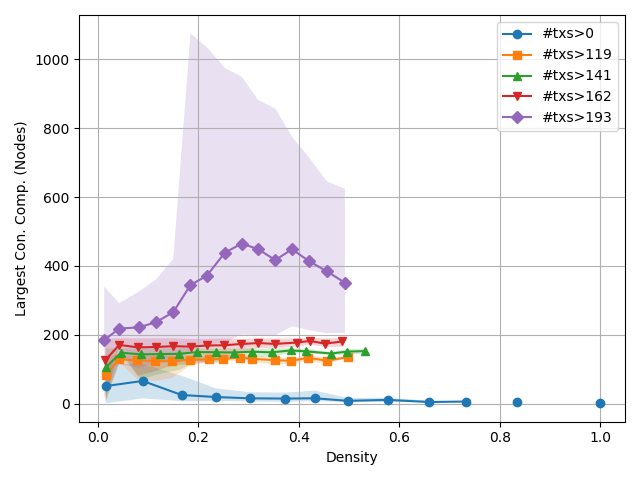}}
\caption{Prestate conflict graph largest connected component by  nodes}
\label{fig:con_comp}
\end{figure}

\begin{figure}
\centerline{\includegraphics[width=\columnwidth]{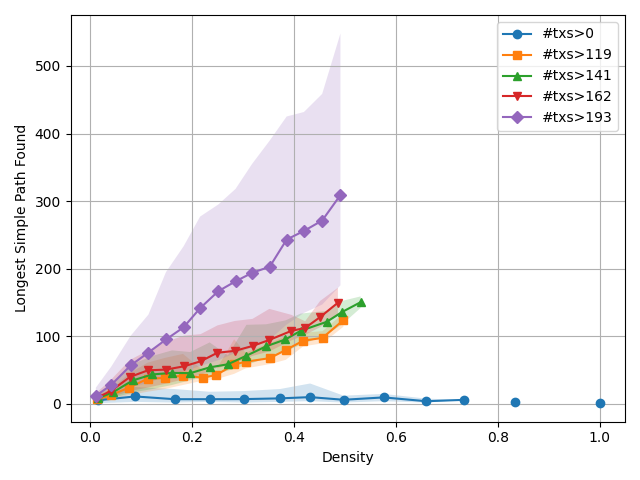}}
\caption{Prestate conflict graph longest simple path found}
\label{fig:long_path}
\end{figure}

\begin{figure}
\centerline{\includegraphics[width=\columnwidth]{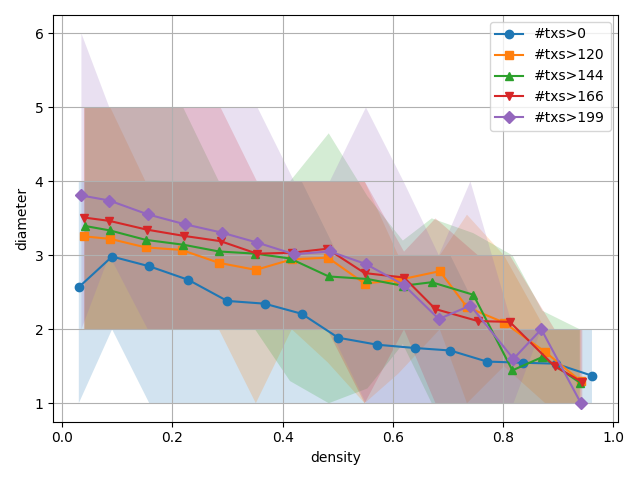}}
\caption{Call conflict graph diameter}
\label{fig:call_diameter}
\end{figure}

\begin{figure}
\centerline{\includegraphics[width=\columnwidth]{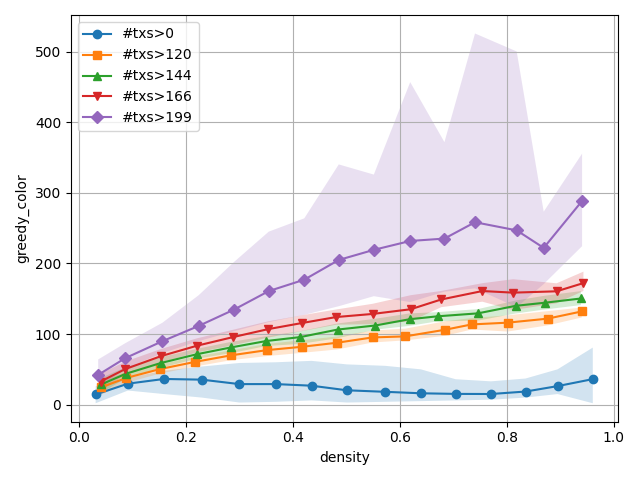}}
\caption{Call conflict graph greedy color}
\label{fig:call_greedy_color}
\end{figure}

\begin{figure*}[!t] 
    \centering
    \subfloat[Lower bound]{
        \includegraphics[width=0.45\linewidth]{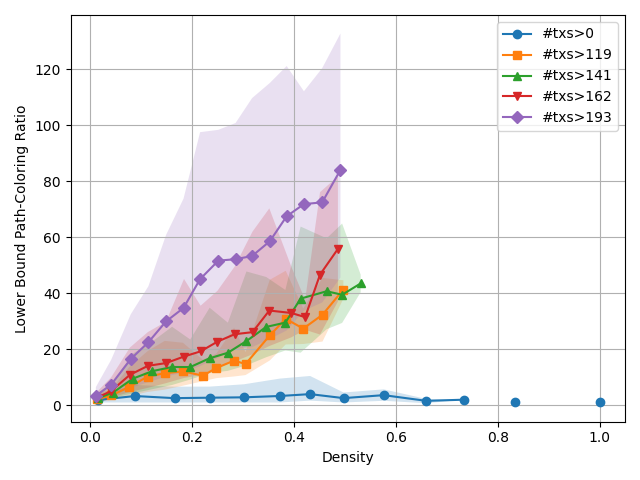}
        \label{fig:lower_bound}
    }
    \subfloat[Upper bound]{
        \includegraphics[width=0.45\linewidth]{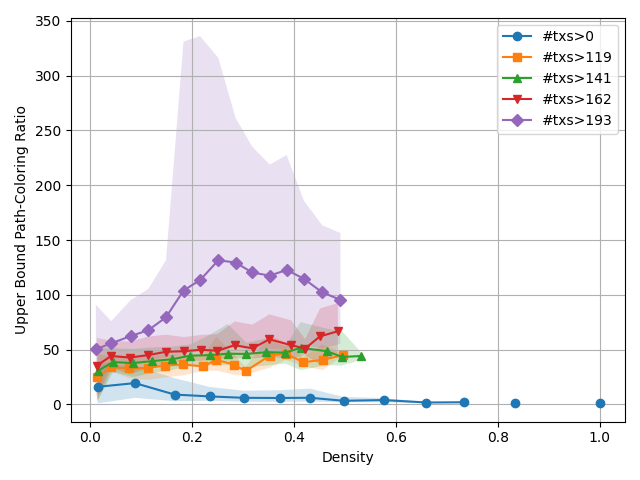}
        \label{fig:upper_bound}
    }
    \caption{Lower and Upper bounds on longest simple path to chromatic number ratio}
    \label{fig:ratio-bounds}
\end{figure*}

\subsubsection{Chromatic Number}

The chromatic number of a graph is bounded between two values: the number of colors used by a greedy coloring algorithm (an upper bound) and the size of the largest clique (a lower bound)\footnote{Exact minimum coloring is an NP-hard problem~\cite{Karp1972}.}. As illustrated in \Cref{fig:greedy_color} and \Cref{fig:clique_number}, these bounds are sufficiently close to each other with no noticeable gap between them.

Interestingly, the chromatic number appears largely unaffected by the graph’s density. This can be explained by star-like structures, where a single transaction conflicts with many others. Such a structure does not increase the chromatic number significantly. The only noticeable exception occurs in the group with the smallest blocks. This group contains empty blocks, which are rare and contain no transactions, resulting in a chromatic number of 0. Aside from these trivial cases, the chromatic number almost always remains in the range of 2 to 5.
As shown in~\cite{ColoringSC}, this gives an excellent potential for~parallelism.

\subsubsection{Longest Simple Path}

Similar reasoning applies to the longest simple path. We consider bounds between the length of a known simple path (a lower bound) and the size of the largest connected component (an upper bound). These metrics are presented in \Cref{fig:con_comp} and \Cref{fig:long_path}.

The largest connected component does not serve as a particularly informative upper bound, especially when the connected components often resemble star graphs. In a star graph, the longest simple path has a length of only 2, regardless of the number of nodes. In contrast, a Monte Carlo approximation of the longest simple path can provide a tighter lower bound.

For small blocks, the longest simple path typically consists of only a few nodes (single-digit lengths). However, larger blocks can have significantly longer paths. At the extreme, the longest paths may exceed 500 nodes and occasionally reach lengths of up to 1200 nodes.

\subsubsection{Longest Simple Path to Chromatic Number ratio}

To estimate the ratio between the longest simple path length and the chromatic number, we use two pairs of metrics. For a conservative lower bound, we divide the Monte Carlo–estimated longest simple path by the greedy coloring number. For a less conservative upper bound, we divide the size of the largest connected component by the clique number.

We present the lower bound in \Cref{fig:lower_bound} and the upper bound in \Cref{fig:upper_bound}.
As the percentage of conflicts increases, the ratio of the longest simple path to the chromatic number also grows. Even the conservative lower bound is virtually always above 2, reaching values near 100 for larger, denser blocks. Notably, the mean of the lower bound is approximately 2.81, and the mean of the upper bound is about 33.01.
This suggests that applying the coloring based technique of~\cite{ColoringSC} for parallelization can be highly effective compared to currently employed methods~\cite{ParBlockchain,sui,sei-2,BlockSTM,monad}.

\subsection{Calltracer Conflict Graph}

We find that the calltracer approach is overly conservative and thus less useful for constructing meaningful conflict graphs. 
Because numerous transactions rely on the same smart contracts, particularly internal contracts with an infrastructural purpose, an excessively high number of collisions occurs.
Consequently, when examining the greedy coloring results for the calltracer approach in \Cref{fig:call_greedy_color}, the number of colors required is nearly equal to the number of nodes in the graph. 
This value is roughly two orders of magnitude larger than that observed with the prestate tracer.

In contrast, the diameter results shown in \Cref{fig:call_diameter} remain relatively modest. With a median diameter around 5, the underlying structure still appears star-like. Thus, although the calltracer yields graphs that are densely interconnected, they retain a fundamentally star-oriented topology.

\section{Discussion}
\label{sec:discussion}

We have found that smart contracts account for a significant portion of all transactions inside Ethereum blocks.
Hence, improving their execution performance is important.
We have also found that the chromatic number of the conflict graphs tend to remain between 2-5, while the average block size is around 150, indicating that Ethereum nodes could greatly benefit from parallel execution of smart contracts.
Further, the high ratio between the longest simple path and the chromatic number of these graphs indicate that Ethereum should consider adopting the coloring approach of~\cite{ColoringSC} to parallelism, as it could provide higher levels of parallelism compared to following an arbitrary block order.

Obviously, our findings capture the current situation for Ethereum.
Should the structure of Ethereum's conflict graphs change in the future, it would impact our conclusions accordingly.
For example, a significant increase in the chromatic number would indicate a significant demise in the potential to benefit from parallelism.
Similarly, a decrease in the ratio between the longest simple path to chromatic number would turn the coloring based approach of~\cite{ColoringSC} less attractive.

We hope that our analysis of the Ethereum call trees, read-sets, writes-ets, and conflict graphs would help in creating more accurate artificial benchmarks.
The latter are important, e.g., for light-weight evaluation of various local execution optimizations.
They are particularly helpful for machine learning based solutions, which require large and diverse training~data.
    
In this work, we have also examined both the prestate tracer and call tracer as methods of extracting conflicts and their respective conflict graphs.
We discovered that while the call tracer approach generated similarly star-like graphs, it inflated the number of conflicts significantly, rendering it useless for our purposes.
The data we collected is available in~\cite{EthStats}\ifdefined\FULLVERSION, the entire set of graphs is in~\cite{full-set}, \fi while the code is open sourced in~\cite{Github}.

\textbf{Acknowledgements.} This work was partially funded by the Israel Science Foundation grant \#3119/21 and a grant from the Ethereum Foundation, as part of the 2025 Academic Grants~Round.

%% file: main.bbl
\begin{thebibliography}{10}
\providecommand{\url}[1]{#1}
\csname url@samestyle\endcsname
\providecommand{\newblock}{\relax}
\providecommand{\bibinfo}[2]{#2}
\providecommand{\BIBentrySTDinterwordspacing}{\spaceskip=0pt\relax}
\providecommand{\BIBentryALTinterwordstretchfactor}{4}
\providecommand{\BIBentryALTinterwordspacing}{\spaceskip=\fontdimen2\font plus
\BIBentryALTinterwordstretchfactor\fontdimen3\font minus \fontdimen4\font\relax}
\providecommand{\BIBforeignlanguage}[2]{{%
\expandafter\ifx\csname l@#1\endcsname\relax
\typeout{** WARNING: IEEEtran.bst: No hyphenation pattern has been}%
\typeout{** loaded for the language `#1'. Using the pattern for}%
\typeout{** the default language instead.}%
\else
\language=\csname l@#1\endcsname
\fi
#2}}
\providecommand{\BIBdecl}{\relax}
\BIBdecl

\bibitem{EthStats}
\BIBentryALTinterwordspacing
D.~D. Biton, ``{Ethereum Statistics},'' 2025, accessed: 2025-03-24. [Online]. Available: \url{https://huggingface.co/datasets/dbiton/EthereumStatistics}
\BIBentrySTDinterwordspacing

\bibitem{Github}
\BIBentryALTinterwordspacing
------, ``Ethgrapher,'' 2025, gitHub repository, accessed: 2025-03-24. [Online]. Available: \url{https://github.com/dbiton/EthGrapher}
\BIBentrySTDinterwordspacing

\bibitem{Schneider1990}
F.~B. Schneider, ``{Implementing Fault-tolerant Services Using the State Machine Approach: A Tutorial},'' \emph{ACM Comput. Surv.}, vol.~22, no.~4, pp. 299--319, Dec. 1990.

\bibitem{diablo}
V.~Gramoli, R.~Guerraoui, A.~Lebedev, C.~Natoli, and G.~Voron, ``{Diablo: A Benchmark Suite for Blockchains},'' in \emph{Proc. of the ACM European Conference on Computer Systems (EuroSys)}, 2023, p. 540–556.

\bibitem{ParBlockchain}
M.~J. Amiri, D.~Agrawal, and A.~El~Abbadi, ``{ParBlockchain: Leveraging Transaction Parallelism in Permissioned Blockchain Systems},'' in \emph{Proc. of the 39th IEEE International Conference on Distributed Computing Systems (ICDCS)}, 2019, pp. 1337--1347.

\bibitem{ConcurrencySC}
T.~Dickerson, P.~Gazzillo, M.~Herlihy, and E.~Koskinen, ``{Adding Concurrency to Smart Contracts},'' in \emph{Proc. of the ACM Symposium on Principles of Distributed Computing (PODC)}, 2017, pp. 303--312.

\bibitem{BlockSTM}
R.~Gelashvili, A.~Spiegelman, Z.~Xiang, G.~Danezis, Z.~Li, D.~Malkhi, Y.~Xia, and R.~Zhou, ``{Block-STM: Scaling Blockchain Execution by Turning Ordering Curse to a Performance Blessing},'' in \emph{Proc. of the 28th ACM SIGPLAN Symposium on Principles and Practice of Parallel Programming (PPoPP)}, 2023, p. 232–244.

\bibitem{KD04}
R.~Kotla and M.~Dahlin, ``{High throughput Byzantine fault tolerance},'' in \emph{Proc. of the IEEE/IFIP International Conference on Dependable Systems and Networks (DSN)}, 2004, pp. 575--584.

\bibitem{EmpSdy-Con}
V.~Saraph and M.~Herlihy, ``{An Empirical Study of Speculative Concurrency in Ethereum Smart Contracts},'' ser. Proc. of Tokenomics, 2019.

\bibitem{Blochchain-DB}
S.~Nathan, C.~Govindarajan, A.~Saraf, M.~Sethi, and P.~Jayachandran, ``{Blockchain Meets Database: Design and Implementation of a Blockchain Relational Database},'' \emph{Proc. of the VLDB Endowment}, vol.~12, no.~11, p. 1539–1552, Jul. 2019.

\bibitem{ColoringSC}
Y.~Hay and R.~Friedman, ``{Batch-Schedule-Execute: On Optimizing Concurrent Deterministic Scheduling for Blockchains},'' in \emph{Proc. of IEEE International Symposium on Reliable Distributed Systems (SRDS)}, Oct. 2024.

\bibitem{YCSB}
B.~F. Cooper, A.~Silberstein, E.~Tam, R.~Ramakrishnan, and R.~Sears, ``{Benchmarking Cloud Serving Systems with YCSB},'' in \emph{Proc. of the ACM Symposium on Cloud Computing (SoCC)}, 2010, p. 143–154.

\bibitem{tpc-c}
\BIBentryALTinterwordspacing
T.~P.~P. Council. (1992) {TPC-C: On-Line Transaction Processing Benchmark}. [Online]. Available: \url{https://www.tpc.org/tpcc/default5.asp}
\BIBentrySTDinterwordspacing

\bibitem{tpc-e}
\BIBentryALTinterwordspacing
------. (2007) {TPC-E: On-Line Transaction Processing Benchmark}. [Online]. Available: \url{https://www.tpc.org/tpce/default5.asp}
\BIBentrySTDinterwordspacing

\bibitem{ethereum}
V.~Buterin, ``{Ethereum White Paper: A Next Generation Smart Contract \& Decentralized Application Platform},'' 2013.

\bibitem{Dissecting-EIP-2930}
L.~Heimbach, Q.~Kniep, Y.~Vonlanthen, R.~Wattenhofer, and P.~Züst, ``{Dissecting the EIP-2930 Optional Access Lists},'' 2023.

\bibitem{VH20}
V.~Saraph and M.~Herlihy, ``{An Empirical Study of Speculative Concurrency in Ethereum Smart Contracts},'' in \emph{International Conference on Blockchain Economics, Security and Protocols (Tokenomics)}, vol.~71.\hskip 1em plus 0.5em minus 0.4em\relax Schloss Dagstuhl -- Leibniz-Zentrum f{\"u}r Informatik, 2020.

\bibitem{COS}
I.~A. Escobar, E.~Alchieri, F.~L. Dotti, and F.~Pedone, ``{Boosting Concurrency in Parallel State Machine Replication},'' in \emph{Proc. of the 20th ACM/IFIP International Middleware Conference}, 2019, p. 228–240.

\bibitem{solana-sealevel}
A.~Yakovenko, ``{Sealevel — Parallel Processing Thousands of Smart Contracts},'' \url{https://medium.com/solana-labs/sealevel-parallel-processing-thousands-of-smart-contracts-d814b378192}.

\bibitem{sui}
S.~Foundation, ``{All About Parallelization},'' \url{https://blog.sui.io/parallelization-explained}.

\bibitem{sei-2}
Sei, ``{Sei v2 - The First Parallelized EVM Blockchain},'' \url{https://blog.sei.io/sei-v2-the-first-parallelized-evm/}.

\bibitem{monad}
M.~Labs, ``{Parallel Execution},'' \url{https://docs.monad.xyz/technical-discussion/execution/parallel-execution}.

\bibitem{DetOverview}
D.~J. Abadi and J.~M. Faleiro, ``{An Overview of Deterministic Database Systems},'' \emph{Communications of the ACM}, vol.~61, no.~9, pp. 78--88, sep 2018.

\bibitem{Aria}
Y.~Lu, X.~Yu, L.~Cao, and S.~Madden, ``{Aria: A Fast and Practical Deterministic OLTP Database},'' \emph{Proc. VLDB Endow.}, vol.~13, no.~12, p. 2047–2060, jul 2020.

\bibitem{caracal}
D.~Qin, A.~D. Brown, and A.~Goel, ``{Caracal: Contention Management with Deterministic Concurrency Control},'' in \emph{Proc. of the ACM Symposium on Operating Systems Principles (SOSP)}, 2021, p. 180–194.

\bibitem{EthPerfAnalysis}
S.~Rouhani and R.~Deters, ``{Performance Analysis of Ethereum Transactions in Private Blockchain},'' in \emph{Proc. of IEEE ICSESS}, 2017, pp. 70--74.

\bibitem{EthInfoProp}
S.~K. Behfar and J.~Crowcroft, ``{Analysis of Information Propagation in Ethereum Network Using Combined Graph Attention Network and Reinforcement Learning to Optimize Network Efficiency and Scalability},'' 2023.

\bibitem{EthGasSize}
T.~Q. Do and M.~T. Ta, ``{Performance Analysis of Ethereum Smart Contracts: A Study on Gas Cost and Block Size Impact},'' in \emph{Proc. of the IEEE Statistical Signal Processing Workshop (SSP)}, 2023, pp. 591--595.

\bibitem{ETHLargeBlock}
P.~Ocheja, M.~Cortes-Goicoechea, T.~Mohandas-Daryanani, B.~Flanagan, H.~Ogata, J.~L. Munoz, and L.~Bautista-Gomez, ``{An Analytical Study of Large Blocks on Ethereum},'' in \emph{Proc. of the ACM Blockchain and Internet of Things Conference (BIOTC)}, 2024, p. 120–127.

\bibitem{XBlock-ETH}
P.~Zheng, Z.~Zheng, and H.~ning Dai, ``{XBlock-ETH: Extracting and Exploring Blockchain Data From Ethereum},'' arXiv 1911.00169, 2019.

\bibitem{RoleReward}
D.~Mancino, A.~Leporati, M.~Viviani, and G.~Denaro, ``{A Role and Reward Analysis in Off-chain Mechanisms for Executing MEV Strategies in Ethereum Proof-of-Stake},'' \emph{ACM Distributed Ledger Technology}, Jun. 2024.

\bibitem{geth_tracing}
\BIBentryALTinterwordspacing
E.~Foundation, ``Built-in tracers in geth,'' 2025, accessed: 2025-03-24. [Online]. Available: \url{https://geth.ethereum.org/docs/developers/evm-tracing/built-in-tracers}
\BIBentrySTDinterwordspacing

\bibitem{dsatur}
D.~Br\'{e}laz, ``{New Methods to Color the Vertices of a Graph},'' \emph{Communications of ACM}, vol.~22, no.~4, pp. 251–--256, Apr. 1979.

\bibitem{Karp1972}
R.~M. Karp, \emph{{Reducibility Among Combinatorial Problems}}.\hskip 1em plus 0.5em minus 0.4em\relax Springer, 1972, pp. 85--103.

\bibitem{gnp-dfs}
S.~Diskin and M.~Krivelevich, ``{On the Performance of the Depth First Search Algorithm in Supercritical Random Graphs},'' arXiv 2111.07345, 2022.

\bibitem{full-set}
\BIBentryALTinterwordspacing
D.~D. Biton. {Additional Graphs}. [Online]. Available: \url{https://bitbucket.org/ethstats/ethstats/}
\BIBentrySTDinterwordspacing

\end{thebibliography}
